\documentclass[12pt]{iopart}

\begin{document}

\title{Nonlocal variables with product states eigenstates}

\author{Berry Groisman\dag ~and Lev Vaidman\dag \ddag
}

\address{\dag  ~School of Physics and Astronomy,
Raymond and Beverly Sackler Faculty of Exact Sciences,
Tel-Aviv University, Tel-Aviv 69978, Israel}

\address{\ddag ~Centre for Quantum Computation,
Department of Physics, University of Oxford,
 Clarendon Laboratory, Parks Road, Oxford OX1 3PU, UK}

\begin{abstract}
An alternative proof for existence of ``quantum nonlocality without
entanglement'', i.e. existence of variables with product-state
eigenstates which cannot be measured locally,  is presented. A simple
``nonlocal'' variable for the case of one-way communication is given
and the limit for its approximate measurability is found.
\end{abstract}

\section{Introduction}
\label{Introduction}
A {\it nonlocal variable} is a property  of a compound quantum
system which cannot be measured using measurements of local
properties only. Aharonov and his collaborators performed an
extensive analysis of  nonlocal variables \cite{aa,nvar,caus}
motivated by the question: to which extent quantum states and
quantum variables have ``physical reality''? Here ``real''
corresponds to ``measurable''. For this analysis, it was crucial
that the measurements of local properties were performed
simultaneously (in some Lorentz frame). The resources were not
constrained: measuring devices included, in particular, entangled
quantum systems. It was found that there are nonlocal variables
which are measurable using only local interactions and prior 
entanglement. In particular, the Bell operator, the
eigenstates of which are four maximally entangled states, is
measurable. On the other hand, it was
proven that there are unmeasurable variables too.

Today, nonlocal variables became an important concept for
practical applications in the field of quantum communication. The
constraint of simultaneity of local measurements is, usually, not relevant
for these considerations, but instead, there are constraints on
resources. The standard  question is: What can be measured
with local measurements and unlimited classical communication?
 It is assumed that the measuring devices do not include
entangled systems, otherwise, the quantum states of the
parts of the composite systems could be all  {\it teleported}
\cite{tele} to one place and then the ``nonlocal'' variable
becomes effectively local. The analysis of the locality of
variables according to this definition was recently performed by
Bennett \etal \cite{bennett} and they found that there are
variables with product states eigenstates which are unmeasurable.

In the present work we suggest an alternative, more simple proof of
the main result of Bennett \etal. We apply our method first to a
similar problem in which only one-way classical communication is
allowed, and at the end, to a  generalization  of the result of Bennett \etal which they  suggested  as a conjecture.

\section {One-way classical communication constraint}\label{4}

 We are looking for a variable of
a composite system consisting of two parts $A$ and $B$ which has product-state
eigenstates $|\Psi_i\rangle=|\phi_i\rangle_a|\psi_i\rangle_b$ and which is not  measurable via local  interactions and
{\it one-way} communication from $A$ to $B$.
We take a  minimal definition of ``measurable'': the measurement has to tell with
certainty if the system is in a particular eigenstate of the
measured variable. There is no requirement that the measurement is
ideal, i.e., that the eigenstates are not changed in the process
of measurement: it might be a demolition measurement. Given that
it identifies all the eigenstates, the linearity of quantum theory
ensures that if the initial state is a superposition of the
eigenstates, then the measurement will yield the outcomes with the
probabilities governed by the quantum theory.

It turns out that there is a very simple example of such a variable.
The nondegenerate eigenstates of this variable are:
\begin{eqnarray}\label{4st}
|\Psi_1\rangle=&|0\rangle_a&|0\rangle_b,\nonumber\\
|\Psi_2\rangle=&|1\rangle_a&|0\rangle_b,\\
|\Psi_3\rangle={\scriptstyle\frac{1}{\sqrt2}}(|0\rangle_a&+|1\rangle_a)~&|1\rangle_b,\nonumber\\
|\Psi_4\rangle={\scriptstyle\frac{1}{\sqrt2}}(|0\rangle_a&-|1\rangle_a)~&|1\rangle_b,\nonumber
\end{eqnarray}
where $|0\rangle_a$,  $|1\rangle_a$ is the basis at $A$ and
$|0\rangle_b$, $|1\rangle_b$ is the basis at $B$.
Let us formulate the problem again. The system is prepared by an
external party in one of the four mutually orthogonal product states
$|\Psi_i\rangle$. The prepared state is unknown to Alice who is
located at $A$ and to Bob who is located at $B$. The aim of the
measurement is to find out in which  initial state the system
has been prepared.  The one-way communication channel is from Alice to Bob.
Bob cannot transmit any information to Alice and cannot act on Alice's
state, therefore, Alice has to start first. Alice performs sequence of
measurements and local operations on her part of the system and gets a
particular outcome $k$. She can also perform her operations step by
step, but there is no principal difference between one step and many
steps  strategy; ``$k$'' signifies the final outcome after
Alice completed all her measurements. Alice reports outcome $k$ to Bob.

Alice's quantum measurement can be described by two stages: at the
first stage the evolution of the quantum state is unitary, at the
second stage a collapse of the quantum state  (real or effective)
occurs. It is enough to consider only the first stage. The unitary
evolution on the Alice's part can be described as:
\begin{equation}
\label{U}
|\phi_i\rangle_a|A\rangle\mapsto\sum_{k=1}^K\alpha_{ik}|w_{ik}\rangle_a,
\end{equation}
 where $|A\rangle$ is the  initial quantum state of Alice's  measuring devices,
 $|w_{ik}\rangle_a$ is the quantum state  of the particle and Alice's
 measurement devices corresponding to  a particular outcome $k$, and the summation  is  over
all possible outcomes.

 There are following relations between possible initial states in the site of Alice:
\begin{eqnarray}\label{inrel}
|\phi_3\rangle=\frac{1}{\sqrt2}(|\phi_1\rangle+|\phi_2\rangle),\nonumber\\
|\phi_4\rangle=\frac{1}{\sqrt2}(|\phi_1\rangle-|\phi_2\rangle).
\end{eqnarray}
The unitary evolution (\ref{U}) keeps these relations:
\begin{eqnarray}\label{super0}
\sum_k\alpha_{3k}|w_{3k}\rangle=\frac{1}{\sqrt2}(\sum_k\alpha_{1k}|w_{1k}\rangle+\sum_k\alpha_{2k}|w_{2k}\rangle),\nonumber\\
\sum_k\alpha_{4k}|w_{4k}\rangle=\frac{1}{\sqrt2}(\sum_k\alpha_{1k}|w_{1k}\rangle-\sum_k\alpha_{2k}|w_{2k}\rangle) .
\end{eqnarray}
  We choose  amplitudes $\alpha_{ik}$ to be
real and nonnegative. This is always possible because the phase can be
included in the definition of $|w_{ik}\rangle$.
Quantum states  $|w_{ik}\rangle$  and $|w_{jk'}\rangle$ with different
$k$ and $k'$ are orthogonal  because they correspond to different outcomes of the measuring devices
(which by definition are macroscopic).
Therefore, the relations (\ref{super0}) hold for each  $k$ separately:
\begin{eqnarray}\label{super}
\alpha_{3k}|w_{3k}\rangle=\frac{1}{\sqrt2}(\alpha_{1k}|w_{1k}\rangle+\alpha_{2k}|w_{2k}\rangle),\nonumber\\
\alpha_{4k}|w_{4k}\rangle=\frac{1}{\sqrt2}(\alpha_{1k}|w_{1k}\rangle-\alpha_{2k}|w_{2k}\rangle).
\end{eqnarray}

 Initially, the states in Alice's site are mutually orthogonal in each pair:  $\langle\phi_1|\phi_2\rangle_a = 0$ and
 $\langle\phi_3|\phi_4\rangle_a =0$. Thus, Alice is able
to distinguish between the states in each pair. It is important
because Bob's corresponding local states are identical, so he cannot distinguish
between the states $|\Psi_1\rangle$
and $|\Psi_2\rangle$ and between the states  $|\Psi_3\rangle$ and
$|\Psi_4\rangle$. Therefore, whatever Alice does,
she must retain distinguishability between the states in each pair.
 This means, that if a particular outcome $k$
might come out for both initial states  $|\Psi_1\rangle$ and
$|\Psi_2\rangle$ (or  $|\Psi_3\rangle$ and $|\Psi_4\rangle$), then,
at every stage, the corresponding quantum states at Alice's site must
be orthogonal. This can be formulated in the following equations:
\begin{eqnarray}
  \label{ort}
\alpha_{1k}\alpha_{2k}  \langle{w_{1k}}|w_{2k}\rangle =0,\nonumber \\
\alpha_{3k}\alpha_{4k}\langle{w_{3k}}|w_{4k}\rangle =0.
\end{eqnarray}
 From
 (\ref{super}) after some manipulation we obtain:
 \begin{eqnarray}\label{super2}
\alpha_{1k}^2=\frac{1}{2}(\alpha_{3k}^2+\alpha_{4k}^2+2\alpha_{3k}\alpha_{4k}\langle{w_{3k}}|w_{4k}\rangle),\nonumber\\
\alpha_{2k}^2=\frac{1}{2}(\alpha_{3k}^2+\alpha_{4k}^2-2\alpha_{3k}\alpha_{4k}\langle{w_{3k}}|w_{4k}\rangle),\\
\alpha_{3k}^2=\frac{1}{2}(\alpha_{1k}^2+\alpha_{2k}^2+2\alpha_{1k}\alpha_{2k}\langle{w_{1k}}|w_{2k}\rangle),\nonumber\\
\alpha_{4k}^2=\frac{1}{2}(\alpha_{1k}^2+\alpha_{2k}^2-2\alpha_{1k}\alpha_{2k}\langle{w_{1k}}|w_{2k}\rangle).\nonumber
\end{eqnarray}
Substituting (\ref{ort}) in (\ref{super2}), we obtain four equations
for $\alpha_{ik}$  which result in the equality
\begin{equation}
 \label{=} \alpha_{1k}=\alpha_{2k}=\alpha_{3k}=\alpha_{4k}.
\end{equation}
Therefore, if $k$ is a possible outcome for one initial state
$|\Psi_i\rangle$, i.e.,  the corresponding coefficient $\alpha_{ik}$ does not vanish,
 then  $k$ is a possible outcome for all initial states. From
(\ref{ort}) it follows that for all such outcomes $k$, the
orthogonality condition holds:
\begin{equation}
  \label{ort1}
  \langle{w_{1k}}|w_{2k}\rangle =0.
\end{equation}
 Substituting (\ref{=}) into
(\ref{super}) we obtain, for each possible outcome $k$, the following
relations
\begin{eqnarray}
|w_{3k}\rangle=\frac{1}{\sqrt2}(|w_{1k}\rangle +|w_{2k}\rangle),\nonumber\\
|w_{4k}\rangle=\frac{1}{\sqrt2}(|w_{1k}\rangle -|w_{2k}\rangle).
\end{eqnarray}
Thus, the evolutions for different initial states of the quantum
state of the system and Alice's measuring devices which ended with
the outcome $k$ are:
\begin{eqnarray}
\label{44}
|\Psi_1\rangle|A\rangle\rightarrow&|w_{1k}\rangle_a&|0\rangle_b,\nonumber\\
|\Psi_2\rangle|A\rangle\rightarrow&|w_{2k}\rangle_a&|0\rangle_b,\\
|\Psi_3\rangle|A\rangle\rightarrow{\scriptstyle\frac{1}{\sqrt2}}&(|w_{1k}\rangle_a+|w_{2k}\rangle_a)&|1\rangle_b,\nonumber\\
|\Psi_4\rangle|A\rangle\rightarrow{\scriptstyle\frac{1}{\sqrt2}}&(|w_{1k}\rangle_a-|w_{2k}\rangle_a)&|1\rangle_b,\nonumber
\end{eqnarray}
 Taking into account (\ref{ort1}), we see that this structure is isomorphic with
the initial structure (with the correspondence
$|0\rangle_a|A\rangle\leftrightarrow|w_{1k}\rangle_a$,
$|1\rangle_a|A\rangle\leftrightarrow|w_{2k}\rangle_a$). Therefore,
we have shown that if there is a constraint on Alice's actions
such that she  cannot lead to a situation in which it is
impossible in principle to distinguish with certainty between
different initial states $|\Psi_i\rangle$, then she cannot make
any progress towards distinguishing the states. Thus, Alice cannot
give Bob any useful information. Bob can perform  operations on
his local part, but he obviously cannot distinguish between the
states $|\Psi_1\rangle$ and $|\Psi_2\rangle$ and between the states
$|\Psi_3\rangle$ and $|\Psi_4\rangle$. This completes the proof of unmeasurability of the variable with nondegenerate eigenstates (\ref{4st}).

Note that this proof is easily generalized for the variable with the
nondegenerate eigenstates
\begin{eqnarray}
\label{4g}
|\Psi_1\rangle=&|0\rangle_a&|0\rangle_b,\nonumber\\
|\Psi_2\rangle=&|1\rangle_a&|0\rangle_b.\\
|\Psi_3\rangle=&(\cos{\theta}|0\rangle_a+\sin{\theta}|1\rangle_a)&|1\rangle_b\nonumber,\\
|\Psi_4\rangle=&(\sin{\theta}|0\rangle_a-\cos{\theta}|1\rangle_a)&|1\rangle_b\nonumber,
\end{eqnarray}
where
$0 <\theta <\frac{\pi}{2}$.

\section{The two-way classical communication constraint}\label{ns}

In this section we will reproduce the result of Bennett \etal \cite{bennett} using the method of the previous section.  We will
prove that certain variables with product-state eigenstates cannot be
measured (even in the above, demolition way) using
local operations and unlimited classical two-way communication.

 The variable which Bennett \etal found
has the following nondegenerate eigenstates:
\begin{eqnarray}
\label{nine}
|\Psi_1\rangle={\scriptstyle\frac{1}{\sqrt2}}&|0\rangle_a&(|0\rangle_b+|1\rangle_b),\nonumber\\
|\Psi_2\rangle={\scriptstyle\frac{1}{\sqrt2}}&|0\rangle_a&(|0\rangle_b-|1\rangle_b),\nonumber\\
|\Psi_3\rangle={\scriptstyle\frac{1}{\sqrt2}}&|2\rangle_a&(|1\rangle_b+|2\rangle_b),\nonumber\\
|\Psi_4\rangle={\scriptstyle\frac{1}{\sqrt2}}&|2\rangle_a&(|1\rangle_b-|2\rangle_b),\nonumber\\
|\Psi_5\rangle={\scriptstyle\frac{1}{\sqrt2}}&(|0\rangle_a+|1\rangle_a)&|2\rangle_b ,\\
|\Psi_6\rangle={\scriptstyle\frac{1}{\sqrt2}}&(|0\rangle_a-|1\rangle_a)&|2\rangle_b ,\nonumber\\
|\Psi_7\rangle={\scriptstyle\frac{1}{\sqrt2}}&(|1\rangle_a+|2\rangle_a)&|0\rangle_b ,\nonumber\\
|\Psi_8\rangle={\scriptstyle\frac{1}{\sqrt2}}&(|1\rangle_a-|2\rangle_a)&|0\rangle_b ,\nonumber\\
|\Psi_9\rangle=&|1\rangle_a&|1\rangle_b ,\nonumber
\end{eqnarray}
where $|0\rangle$, $|1\rangle$, and
$|2\rangle$ are  local bases in Alice's and Bob's sites.

In our approach, in contrast with the original proof, we will show
that if we impose the constraint of not allowing for any
probability to fail in the measurement, i.e. of reaching a state
in which it is in principle
   impossible
 to distinguish with certainty between different initial
states $|\Psi_i\rangle$, then Alice and Bob cannot make any
progress towards completing the measurement. We note that even if
the two-way communication is allowed, one party has to start.
Since they have only classical channel, a measurement which ends
up with a particular outcome has to be performed in one of the
sites. Thus, constructing the proof similar to that of the
previous section, but for the variable with the eigenstates
(\ref{nine}), is enough.

Since the eigenstates (\ref{nine}) have  a symmetry between $A$ and $B$,
we can assume without loosing generality that  the  first step is
performed by Alice who performs the measurement with possible
outcomes $k$. The
 unitary evolution
on the Alice's part can be described as:
\begin{equation}
\label{evol}
|\phi_i\rangle_a|A\rangle\mapsto\sum_{k}\alpha_{ik}|w_{ik}\rangle_a .
\end{equation}
From $|\phi_1\rangle =|\phi_2\rangle$ and $|\phi_3\rangle
=|\phi_4\rangle$, we immediately obtain:
\begin{eqnarray}
 \label{1234}
 \alpha_{1k} =\alpha_{2k},~~~~~~~\alpha_{3k}=\alpha_{4k},\nonumber \\
|w_{1k}\rangle =|w_{2k}\rangle, ~~~~~~~~
|w_{3k}\rangle=|w_{4k}\rangle .
\end{eqnarray}
The evolution (\ref{evol}) should keep the relations between initial states,
and since all states $|w_{ik}\rangle_a$ with different $k$ must be
orthogonal, the same relations hold for each individual possible
$k$:

\break

\begin{eqnarray}
\label{supernine}
\alpha_{1k}|w_{1k}\rangle=\frac{1}{\sqrt2}(\alpha_{5k}|w_{5k}\rangle+\alpha_{6k}|w_{6k}\rangle),\nonumber\\
\alpha_{3k}|w_{3k}\rangle=\frac{1}{\sqrt2}(\alpha_{7k}|w_{7k}\rangle-\alpha_{8k}|w_{8k}\rangle),\nonumber\\
\alpha_{5k}|w_{5k}\rangle=\frac{1}{\sqrt2}(\alpha_{1k}|w_{1k}\rangle+\alpha_{9k}|w_{9k}\rangle),\nonumber\\
\alpha_{6k}|w_{6k}\rangle=\frac{1}{\sqrt2}(\alpha_{1k}|w_{1k}\rangle-\alpha_{9k}|w_{9k}\rangle),\\
\alpha_{7k}|w_{7k}\rangle=\frac{1}{\sqrt2}(\alpha_{9k}|w_{9k}\rangle+\alpha_{3k}|w_{3k}\rangle),\nonumber\\
\alpha_{8k}|w_{8k}\rangle=\frac{1}{\sqrt2}(\alpha_{9k}|w_{9k}\rangle-\alpha_{3k}|w_{3k}\rangle).\nonumber
\end{eqnarray}

Bob, obviously, cannot distinguish between states $|\Psi_5\rangle$ and
$|\Psi_6\rangle$ and between states $|\Psi_7\rangle$ and
$|\Psi_8\rangle$. He also might fail to distinguish between $|\Psi_1\rangle$ and 
$|\Psi_9\rangle$ and between  $|\Psi_3\rangle$ and 
$|\Psi_9\rangle$ because  the
states $|\Psi_1\rangle$, $|\Psi_3\rangle$ are nonorthogonal to
$|\Psi_9\rangle$ on Bob's side. Therefore, Alice should be able to distinguish between these pairs, i.e.,  for all possible outcomes $k$ the
following  conditions must be kept at Alice's site:
\begin{eqnarray}
\label{ortnine}
\alpha_{5k}\alpha_{6k}\langle{w_{5k}}|w_{6k}\rangle =0,\nonumber
\\ \alpha_{7k}\alpha_{8k}\langle{w_{7k}}|w_{8k}\rangle =0,\nonumber\\
\alpha_{1k}\alpha_{9k}\langle{w_{1k}}|w_{9k}\rangle =0,\\
\alpha_{3k}\alpha_{9k}\langle{w_{3k}}|w_{9k}\rangle =0.\nonumber
\end{eqnarray}
After some straightforward algebraic manipulations,
(\ref{supernine}) and  (\ref{ortnine}) yield that all nine coefficients $\alpha_{ik}$ are  equal, and that $|w_{1k}\rangle_a$, $|w_{3k}\rangle_a$, and $|w_{9k}\rangle_a$ are mutually orthogonal.
 Therefore,  the evolutions for different
initial states of the quantum state of the system and Alice's
measuring devices ended with the outcome $k$ is:
\begin{eqnarray}
|\Psi_1\rangle|A\rangle\rightarrow{\scriptstyle\frac{1}{\sqrt2}}&|w_{1k}\rangle_a&(|0\rangle_b+|1\rangle_b),\nonumber\\
|\Psi_2\rangle|A\rangle\rightarrow{\scriptstyle\frac{1}{\sqrt2}}&|w_{1k}\rangle_a&(|0\rangle_b-|1\rangle_b),\nonumber\\
|\Psi_3\rangle|A\rangle\rightarrow{\scriptstyle\frac{1}{\sqrt2}}&|w_{3k}\rangle_a&(|1\rangle_b+|2\rangle_b),\nonumber\\
|\Psi_4\rangle|A\rangle\rightarrow{\scriptstyle\frac{1}{\sqrt2}}&|w_{3k}\rangle_a&(|1\rangle_b-|2\rangle_b),\nonumber\\
|\Psi_5\rangle|A\rangle\rightarrow{\scriptstyle\frac{1}{\sqrt2}}&(|w_{1k}\rangle_a+|w_{9k}\rangle_a)&|2\rangle_b,\\
|\Psi_6\rangle|A\rangle\rightarrow{\scriptstyle\frac{1}{\sqrt2}}&(|w_{1k}\rangle_a-|w_{9k}\rangle_a)&|2\rangle_b,\nonumber\\
|\Psi_7\rangle|A\rangle\rightarrow{\scriptstyle\frac{1}{\sqrt2}}&(|w_{9k}\rangle_a+|w_{3k}\rangle_a)&|0\rangle_b,\nonumber\\
|\Psi_8\rangle|A\rangle\rightarrow{\scriptstyle\frac{1}{\sqrt2}}&(|w_{9k}\rangle_a-|w_{3k}\rangle_a)&|0\rangle_b,\nonumber\\
|\Psi_9\rangle|A\rangle\rightarrow&|w_{9k}\rangle_a&|1\rangle_b.\nonumber
\end{eqnarray}
 This structure is isomorphic with
the structure of the initial state (with the correspondence
$|0\rangle_a|A\rangle\leftrightarrow|w_{1k}\rangle_a,|1\rangle_a|A\rangle\leftrightarrow|w_{9k}\rangle_a$,
and $|2\rangle_a|A\rangle\leftrightarrow|w_{3k}\rangle_a$).
Therefore, we have shown that if there is a constraint on Alice's
actions such that she cannot lead to a situation in which it is
impossible in principle to distinguish with certainty between
different initial states $|\Psi_i\rangle$, then she cannot make
any progress towards distinguishing the states.
 Thus, Alice cannot
give Bob any useful information.
If Alice's operation (the
first round) yields no  progress towards the solution of the
problem, then all following rounds cannot change the situation
too.

 We can apply this method to prove unmeasurability of a more general variable  suggested in the Bennett's paper as a conjecture. The
 set of nondegenerate eigenstates  of this variable is:
  \begin{eqnarray}
    \label{nineg}
|\Psi_1\rangle=&|0\rangle_a&({\cos{\eta}}|0\rangle_b+\sin{\eta}|1\rangle_b),\nonumber\\
|\Psi_2\rangle=&|0\rangle_a&({\sin{\eta}}|0\rangle_b-\cos{\eta}|1\rangle_b),\nonumber\\
|\Psi_3\rangle=&|2\rangle_a&({\cos{\xi}}|1\rangle_b+\sin{\xi}|2\rangle_b),\nonumber\\
|\Psi_4\rangle=&|2\rangle_a&({\sin{\xi}}|1\rangle_b-\cos{\xi}|1\rangle_b),\nonumber\\
|\Psi_5\rangle=&({\cos{\theta}}|0\rangle_a+\sin{\theta}|1\rangle_a)&|2\rangle_b,\\
|\Psi_6\rangle=&({\sin{\theta}}|0\rangle_a-\cos{\theta}|1\rangle_a)&|2\rangle_b,\nonumber\\
|\Psi_7\rangle=&({\cos{\gamma}}|1\rangle_a+\sin{\gamma}|2\rangle_a)&|0\rangle_b,\nonumber\\
|\Psi_8\rangle=&({\sin{\gamma}}|1\rangle_a-\cos{\gamma}|2\rangle_a)&|0\rangle_b,\nonumber\\
|\Psi_9\rangle=&|1\rangle_a&|1\rangle_b ,\nonumber
 \end{eqnarray}
where all angles $\eta,\xi,\theta,\gamma$ are strictly inside the
interval $(0,{\pi\over2})$. Indeed, considering Alice to make the
first step and following the arguments above, we obtain again relations (\ref{1234}) and also the following relations:
\begin{eqnarray}
\alpha_{1k}^2-\alpha_{9k}^2=\cos{2\theta}(\alpha_{5k}^2-\alpha_{6k}^2),\nonumber\\
\alpha_{5k}^2-\alpha_{6k}^2=\cos{2\theta}(\alpha_{1k}^2-\alpha_{9k}^2),\nonumber\\
\alpha_{9k}^2-\alpha_{3k}^2=\cos{2\gamma}(\alpha_{7k}^2-\alpha_{8k}^2),\\
\alpha_{7k}^2-\alpha_{8k}^2=\cos{2\gamma}(\alpha_{9k}^2-\alpha_{3k}^2).\nonumber
\end{eqnarray}
The only solution of these equations is that all coefficients  $\alpha_{ik}$ are
equal, i.e., that Alice cannot make any progress if she makes the
first step. Similar equations are obtained if Bob is to make the
first step, so  he cannot
make any progress either.

 \section{Optimal local estimation measurements: the case of one-way communication}

 In the previous sections we proved that 100\% reliable
measurements of certain variables are impossible. Let us show now,
for the case of the one-way communication, that not just the ideal
case is impossible but also that it is impossible to get close to
it. To show this we will relax both the requirement of 100\% success
and the requirement of 100\% reliability, and will ask the
following question: What is the optimal measurement which can get
the best guess of the prepared state? The ``best'' means that on
average we obtain maximal probability for the correct guess.
If the maximal probability does not approach 1, it means that it is
impossible to construct a protocol in which success and reliability
will approach 100\%.

 We
consider again the situation after Alice completes the measurements
at her site.
The equations (\ref{U}-\ref{super},\ref{super2}) still hold, but the orthogonality
conditions ({\ref{ort},{\ref{ort1}) are
  not imposed.
 Alice has to distinguish between the states $|\Psi_1\rangle$ and
$|\Psi_2\rangle$ and between the states $|\Psi_3\rangle$ and
$|\Psi_4\rangle$. For each $k$ she makes her guess according to
the maximal coefficient $\alpha_{ik}$ in each pair $\alpha_{1k},
\alpha_{2k}$ and $\alpha_{3k},
 \alpha_{4k}$. Bob distinguishes between the pairs with 100\%
 efficiency, therefore, for a given $k$, on average according to the
 initial state, the probability for the correct guess is
 \begin{equation}
   \label{prob1}
   p ={1\over 2} {{\max \{ \alpha_{1k}^{2},
\alpha_{2k}^{2}\} } \over{\alpha_{1k}^{2} +
   \alpha_{2k}^{2}}} + {1\over 2} {{\max\{\alpha_{3k}^{2}, \alpha_{4k}^{2}\}} \over{
   \alpha_{3k}^{2}+ \alpha_{4k}^{2}}} .
\end{equation}
Our task is to find the strategy for Alice such that, on average
on all outcomes $k$, the probability $p$ will become maximal. As
before, since the constraints are separate for each $k$, we just
have to look for a maximum for a particular $k$.
 From (\ref{super2}) we obtain:
 \begin{equation}
 \alpha_{1k}^2+\alpha_{2k}^2=\alpha_{3k}^2+\alpha_{4k}^2 .
 \end{equation}
  Without loosing generality we can assume that
 $\alpha_{1k} > \alpha_{2k}$ and $\alpha_{3k}
> \alpha_{4k}$. Let us define parameters $\gamma$, $\epsilon$ and $\delta$:
   \begin{eqnarray}\label{dd}
      \alpha_{1k}^2=\gamma(1+\epsilon),\nonumber\\
      \alpha_{2k}^2=\gamma(1-\epsilon),\nonumber\\
      \alpha_{3k}^2=\gamma(1+\delta),\\
      \alpha_{4k}^2=\gamma(1-\delta).\nonumber
   \end{eqnarray}
Then $p$, the probability  we have to maximize, becomes
 \begin{equation}
   \label{prob2}
   p ={{2+\epsilon +\delta}\over 4}.
 \end{equation}
From (\ref{super2}) and (\ref{dd}) we obtain
\begin{equation}
\label{p} p= {1\over4}(2+\epsilon+\sqrt{1-\epsilon^2}\langle{w_{1k}}|w_{2k}\rangle) .
 \end{equation}
 The maximum value of $p$ is obtained when
 $\langle{w_{1k}}|w_{2k}\rangle =1$. Then, the optimization on different
 values of $\epsilon$ yields maximum for
 $\epsilon=\frac{1}{\sqrt2}$, and thus, the probability for the
 correct guess of the prepared state is not more than
\begin{equation}
\label{pmax}
 p_{max}=\frac{1}{2}+\frac{1}{2\sqrt{2}}.
\end{equation}
 This bound  is, in fact, tight, since it can be realized
via the measurement in the basis
\begin{eqnarray}
|\chi_1\rangle_a=\sin(\frac{\pi}{8})|0\rangle_a+\cos(\frac{\pi}{8})|1\rangle_a ,
\nonumber \\
|\chi_2\rangle_a=\cos(\frac{\pi}{8})|0\rangle_a-\sin(\frac{\pi}{8})|1\rangle_a .
\end{eqnarray}
 If the state $|\chi_1\rangle$ is obtained, then the team announces that
 its guess is $|\Psi_2\rangle$ or $|\Psi_3\rangle$,  according to the
 results of Bob, $|0\rangle_b$ or $|1\rangle_b$. If the state
 $|\chi_2\rangle$ is obtained, then the team announces
 $|\Psi_1\rangle$ or $|\Psi_4\rangle$ according to the
 results of Bob.
It can be seen from a straightforward trigonometric manipulations that the probability for the correct result is indeed $p_{max}$:
\begin{equation}
\label{ptight}
 |\langle \phi_2|\chi_1\rangle|^2=|\langle \phi_3|\chi_1\rangle|^2 =|\langle \phi_1|\chi_2\rangle|^2 =|\langle \phi_4|\chi_2\rangle|^2 =\frac{1}{2}+\frac{1}{2\sqrt{2}}.
\end{equation}
Note that the probability of the successful guess is the same for all initial states  $|\Psi_i\rangle$.

\section{Conclusions}

In this paper we have discussed measurability of variables of a
composite system consisting of two separated parts. The variables
we have considered  have nondegenerate product-state eigenstates.
We have shown that assuming one-way classical communication and
local interactions, there is a simple example of unmeasurable
variable of this kind with eigenstates given by (\ref{4st}).  We
have simplified the proof and made certain generalization of the
results by Bennett \etal  regarding measurability of such
variables when the two-way classical communication is allowed.

It was shown before \cite{caus} that the variable with product-states
eigenstates (\ref{4st}) is unmeasurable. However, this proof was under
different assumptions. The main difference is that the proof was only
for unmeasurability of ideal (nondemolition) measurements. In this
case it was easy to show that measurability leads to superluminal
signaling and this was the proof that it is impossible. The current
work considers more difficult question of possibility of  ``demolition'' measurements
in which it is not required that the eigenstates are unchanged during
the measurement. Under certain constraints regarding allowed
operations,  it has been shown that a variable with an entangled (but not maximally entangled) eigenstate
 cannot be measured  even in a demolition way \cite{aa,nvar}.  When
the constraints were removed \cite{caus}, the unmeasurability was
not true anymore, but it was shown that measurement of such a
variable invariably erases relevant local information. It seems
that more results can be obtained in this framework. In
particular, a recently developed formalism of {\it semicausal}
operations seems very promising as it already led to useful
results in quantum communication \cite{BGNP}.

Just before the completion of this work, an actual experiment with the
eigenstates (\ref{nine}) has been proposed by Carollo \etal
\cite{CPSZ}.
An experiment with nine eigenstates (\ref{nine}) is significantly more
difficult than an experiment with four states (\ref{4st}). The
impossibility to distinguish the states (\ref{4st}) with local
measurements and the one-way classical communication represents the same
basic feature as the impossibility to distinguish the states
(\ref{nine}) with the two-way communication channel. Thus, we suggest to
modify this experimental proposal   for the case of
four states and to start with this easier experiment.

Carollo \etal also proved that even if the nine eigenstates are
available locally, they cannot be discriminated using linear optical
elements. It means that even the existence of prior entanglement does
not allow reliable discrimination of the nine bipartite states
(\ref{nine}) with linear elements, and not just because the
teleportation cannot be performed \cite{VY,FIN}. The question of
reliable discrimination of the four bipartite states (\ref{4st}) with
linear optics, one-way communication, and prior entanglement remains
open.

\ack

It is a pleasure to thank Noam Erez, Christopher Fuchs, Lior
Goldenberg, Tal Mor, Christoph Simon, Stephen Wiesner, and Nadav Yoran for helpful
discussions.  This research was supported in part by grant 471/98 of
the Basic Research Foundation (administered by the Israel Academy of
Sciences and Humanities) and the EPSRC grant GR/N33058.

\end{document}